# Multidimensional Nature of Molecular Organic Conductors Revealed by Angular Magnetoresistance Oscillations

*Pashupati Dhakal,[a,b] Harukazu Yoshino,[c] Jeong-Il Oh,[a] Koichi Kikuchi[d] and Michael J. Naughton[a,*]*

[a] *Department of Physics, Boston College, Chestnut Hill, MA 02467 USA*

[b] *SRF Institute, Jefferson Lab, Newport News, VA 23606 USA*

[c] *Graduate School of Science, Osaka City University, Osaka 558-8585 Japan*

[d] *Graduate School of Science and Engineering, Tokyo Metropolitan University, Tokyo 192-0397 Japan*

**Abstract**

Angle-dependent magnetoresistance experiments on organic conductors exhibit a wide range of angular oscillations associated with the dimensionality and symmetry of the crystal structure and electron energy dispersion. In particular, characteristics associated with 1, 2, and 3-dimensional electronic motion are separately revealed when a sample is rotated through different crystal planes in a magnetic field. Originally discovered in the TMTSF-based conductors, these effects are particularly pronounced in the related system $(DMET)_2I_3$. Here, experimental and computational results for magnetoresistance oscillations in this material, over a wide range of magnetic field orientations, are presented in such a manner as to uniquely highlight this multidimensional behavior. The calculations employ the Boltzmann transport equation that incorporates the system's triclinic crystal structure, which allows for accurate estimates of the transfer integrals along the crystallographic axes, verifying the 1D, 2D and 3D nature of $(DMET)_2I_3$, as well as crossovers between dimensions in the electronic behavior.

*Keywords: AMRO, $(DMET)_2I_3$, Molecular Organic Conductors*

---

[*] Corresponding author at: Department of Physics, Boston College, Chestnut Hill, MA 02467 USA. Tel.:+1-617-552-3598; Email: naughton@bc.edu

1. **Introduction**

Crystalline molecular organic conductors have attracted much interest in recent years because of their reduced dimensionality and relatively large electron correlations, resulting in exotic physical properties. In particular, the quasi-one dimensional (Q1D) organic conductors, among the more interesting materials in all of condensed matter physics, exhibit a wide variety of electronic and magnetic ground states [1,2]. These materials comprise 1D-like chains along a crystallographic *x*-direction which are moderately coupled along the transverse *y*-direction, giving rise to anisotropic 2D sheets, rather like graphene but with in-plane anisotropy. These sheets are themselves weakly coupled along the crystallographic *z*-direction (akin to graphite). Unlike in like these *omnis* carbon cousins, the coupling between the chains and sheets is such that the hopping integrals $t_i$ are highly anisotropic, with $t_x \gg t_y \gg t_z$, yielding $\sigma_x \gg \sigma_y \gg \sigma_z$, where conductivity components vary as $\sigma_i \sim t_i^2$. Due to the weak interchain and interplane coupling, the Fermi surface contains a pair of warped 1D sheets which support only open orbits in a magnetic field, such that Landau quantization is absent [1,2]. Nonetheless, the electrical conductivity shows large oscillatory phenomena for magnetic field rotated in different crystalline planes. Several related types of so-called angular magnetoresistance oscillations (AMRO) have been observed in these Q1D conductors, as well as in quasi-two-dimensional ones [1-3]. In Q1D, when the magnetic field is rotated about the orthogonal axes, $a//x$, $b'//y$ and $c^*//z$, seemingly different kinds of AMRO are observed, to wit: Lebed magic angle (LMA) resonances [4,5,6,7,8] for field rotated in the plane perpendicular to the one-dimensional chains (*i.e.* about *x*), and Danner-Kang-Chaikin (DKC) oscillations [9] and the Yoshino angular effect (YAE) [10,11] for rotations about *y* and *z*, respectively. In addition, more complex oscillations are observed, initially by Lee and Naughton (LN), for magnetic field rotated in arbitrary, out-of-plane directions [12,13,14]. The LMA and LN oscillations have recently been understood as arising from the same physics [15], such that the moniker LNL has been applied [16]. All these AMRO effects are direct manifestations of the *multi*-dimensionality and high electronic and crystal anisotropies of Q1D conductors.

The AMRO effects have been observed in many such Q1D materials, and their fundamental origins have been intensively investigated over the last two decades. The field essentially was started by Lebed's first theoretical prediction [4] of a novel angle dependence to the threshold field for a Fermi surface nesting induced by a magnetic field, leading to a series of field-induced spin density wave transitions. Burlachkov, Gor'kov and Lebed [17], as well as Lebed and Bak [18], later showed that, in a magnetic field tilted from the normal to the layers, electron motion in the plane is quasi-periodic, and a type of low dimensional limit is reached, *i.e.* for the fields for which the effective frequency to cross the Brillouin zone in the *z*-direction, $\omega_z = ec^* B_y v_F / \hbar$ (where $c^*$ is the interlayer distance) exceeds that associated with the interlayer bandwidth, $\omega_z \geq 4t_z/\hbar$. This corresponds to the limit where the amplitude of electron motion along the interplane *z*-direction becomes smaller than the interlayer distance parameterized by the lattice distance. Above this field, the layers become decoupled (2D), and the electronic dimensionality is reduced sufficiently enough to induce a 1D Peierls-type instability, aided by the intrinsic in-plane anisotropy.

While this scenario may not correspond to experimental reality, it did instigate experimentalists and theorists to take a closer look at the materials and their behavior in tilted magnetic fields. In such a situation, at certain angles (since called magic angles), the periods of electron orbits along the $k_y$ and $k_z$ directions on the open FS sheets become commensurate. Osada, Kagoshima and Miura showed that electrons in Q1D systems can have non-zero average velocity *along* the field direction at those special angles, leading to increase in interlayer conductivity [19]. As a result, when the magnetic field is rotated from the *y* to the *z*-axis, a series of minima in interlayer resistance is expected at those angles. The first experimental evidence of such effect features in magnetoresistance at certain magic angles was observed in the Bechgaard salt (TMTSF)$_2$ClO$_4$ [5,6,7]. Similar effects have since been found in several other Q1D organic conductors, such as (TMTSF)$_2$PF$_6$ [8,20], (TMTSF)$_2$ReO$_4$ [21], (DMET-TSeF)$_2$X (X=AuCl$_2$, AuI$_2$, I$_3$) [22, 23,24], (DMET)$_2$CuCl$_2$ [25],(BEDT-TTF) (TCNQ) [26], as well as in the material discussed here, (DMET)$_2$I$_3$ [16].

Several theoretical models [11,12, 19, 27-34] have been put forth to explain competing 1D-2D-3D effects seen in interlayer AMRO in Q1D materials, based on the semiclassical Boltzmann transport equation, [11, 12, 19, 28] the quantum mechanical Kubo formalism, [15, 29, 30, 31] a phase coherence of interlayer electron tunneling scheme [32, 33, 34]. In this paper, we show the multidimensional nature of Q1D conductors, via detailed interlayer magnetoresistance measurements on (DMET)$_2$I$_3$, for all angular orientation of magnetic field, as well as simulations of the same via numerical calculations employing the appropriate (triclinic) crystal structure. The simulations reveal all the experimentally observed AMRO effects and, presented in polar plots, clearly point out the regimes where different dimensions dominate.

## 2. (DMET)$_2$I$_3$

The non-centrosymmetric molecule dimethyl(ethylenedithio)diselenadithiafulvalene (DMET) is formed by combining half of a tetramethyl-tetraselenafulvalene (TMTSF) molecule and half of a bis-ethylenedithio-tetrathiafulvalene (BEDT-TTF) (a.k.a. ET) molecule. (DMET)$_2$I$_3$ [35] shares many similarities with the TMTSF and other Q1D system, including Q1D AMRO effects, field-induced spin density waves (FISDW), superconductivity, and a triclinic crystal structure. In this symmetry, the orthogonal set (*x, y, z*) is represented by (*b, a', c$^*$*), based on the lattice parameters *b, a* and *c* as shown in Fig. 1(a). As mentioned, several experimental works revealing the various AMRO effects have been completed in $(TMTSF)_2X$. $(DMET)_2I_3$ is probably a better material for the study these Q1D effects, as all important effects occur at ambient pressure.

## 3. Experiment

Single crystals of $(DMET)_2I_3$ were synthesized by conventional chemical oxidation [36]. Experiments were carried out in a dilution refrigerator and split-coil superconducting magnet with a two axis rotator system that covers $4\pi$ steradians of ($\theta, \phi$) angle space. An *ex situ* goniometer gives a complete rotation of the cryostat about refrigerator, while an *in situ* rotator allows the sample stage to rotate about an axis in a plane normal to the refrigerator. For this work, we used a split-coil magnet with maximum field of 11 T at 4.2K, with a 40 mm access gap, and two rotators that include a commercial goniometer (Huber model 420) and a home-built *in situ* rotator. The sample holder was made of OFHC copper, proving a strong thermal link between the mixing chamber and the sample (in vacuum) stage. Two $(DMET)_2I_3$ samples were mounted on the stage, along with three temperature sensors, a LakeShore Cernox CX-1030-SD sensor for temperatures above 2 K, and two $RuO_2$ sensors for low temperatures. One $RuO_2$ sensor was resident on the sample stage and one on the mixing chamber (in far reduced *B*-field), both previously calibrated at zero magnetic field with respect to an Oxford Instruments-supplied $RuO_2$ mixing chamber thermometer as well as nuclear orientation thermometry.

The dilution refrigerator was placed on the goniometer, providing a complete 360° rotation and very accurate rotation of the refrigerator ($\delta\phi$= 0.0025°), allowing us to accurately align our sample with respect to the magnetic field. To avoid any restriction of rotation of the refrigerator, a turntable was designed such that, when the *ex situ* goniometer is rotated in one direction, this turntable allows the cryogenic dewar to simultaneously rotate in the opposite direction, using a third stepper motor. That is, when the sample (in the dilution refrigerator) rotation reached a few degrees CW in fine steps, the dewar + magnet + dilution refrigerator combination was grossly rotated a few degrees CCW, keeping refrigerator nominally static in the laboratory frame. While rotations in a magnetic field can produce temperature fluctuations during the measurements, we found that while rotating the refrigerator using the goniometer (*ex-situ* rotation), there was only a small fluctuation in temperature (<5 mK) that rapidly recovered (seconds). However, the *in situ* rotator can easily cause excessive sample heating at low temperature due to friction, since the sample was directly mounted on the rotator. In order to minimize frictional heat, a Kevlar-based fiber string driven *in situ* rotator was employed instead of a gear-driven rotator for this work, with resulting temperature variations and recovery times ultimately comparable to the external result above.

Two samples with dimensions ~0.5 × 0.3 × 0.15 mm$^3$ were mounted onto a rotating sample stage, contacted with 12 μm diameter gold wires held by graphite paste. The standard four probe measurement technique was employed with RMS currents of 1 μA to monitor the interlayer resistance *($R_{zz}$)* using Stanford Research Systems model 830 lock-in amplifiers. The room temperature resistivities of the samples were ~200 μΩ–cm, indicating good metallicity. Conventional metallic behavior was seen upon cooling, with a superconducting transition in both samples (midpoints) at $T_c$ ~ 0.58 K.

We report data at 100 mK and 9 T for one sample, as the results are qualitatively the same for both. This field is below the minimum of the threshold field for FISDW formation, ensuring the system is always in the normal metal state (field too small for FISDW and too large for superconductivity). This contrasts with most previous AMRO studies in TMTSF systems, where complicated mixtures of metallic and SDW states occurred in measured magnetic fields, tending to complicate interpretation of data.

## 4. Triclinic Calculation

In the semiclassical picture, the motion of an election is described by the equation of motion as $\hbar d\vec{k}/dt = -e\vec{v} \times \vec{B}$, where $\vec{v} = \partial E(\vec{k})/\hbar \partial \vec{k}$. The Boltzmann transport equation is a semiclassical approach to calculate the carrier transport in crystalline metals. The expression for the magnetoconductivity tensor $\bar{\sigma}_{ij}$, under the relaxation time approximation [37] is given by

$$\bar{\sigma}_{ij} = -\frac{2e^2}{V} \sum_k (df/dE) v_i(k,0) \int_{-\infty}^{0} v_j(k,t) e^{t/\tau} dt, \qquad (1)$$

where $e$ = electronic charge, $V$ = sample volume, $f$ = Fermi distribution function, $E$ = electron energy, $v_i$ = i$^{th}$ component of the carrier velocity, $k$ = electron wave vector, $t$ = time, and $\tau$ = relaxation time, respectively, with $\tau$ assumed to be $k$-independent. The carrier velocity can be calculated based on a tight binding energy dispersion,
$E = -2t_b \cos k_b b - 2t_a \cos k_a a - 2t_c \cos k_c c$, where $b$, $a$ and $c$, are lattice parameters and $t_b$, $t_a$ and $t_c$ are intermolecular transfer integrals between neighboring sites along $b$, $a$ and $c$, respectively. We calculate the interlayer magnetoresistance using $t_b:t_a:t_c$ = 300:30:1 [38] and $\tau = 10^{-14}$ sec. It is found that this value of $\tau$ gives clear AMRO structures with the magnitude of $B$ that we can achieve in the laboratory. The velocities $v_b$, $v_a$, and $v_c$ along triclinic crystal axes are calculated and transformed into those along the Cartesian $x$, $y$ and $z$-axes using matrix transformations [39]. Once the Lorentz forces and resulting changes in wave vector during a short period are calculated in the Cartesian system, new wave vectors are converted to the triclinic system by using an inverse matrix transformation. These wave vectors give a new Fermi velocity from the dispersion relation and this procedure is repeated until the integration above converges.

To acquire computational simulation results with sufficient precision, the first Brillouin zone is divided into a grid of $128^3 \sim 2\times10^6$ sites. Time integrations were performed with $\Delta t = 10^{-16}$ s steps (*i.e.*, $\tau/100$), sufficiently accurate for the fields employed. For very high fields ($B > 30$T), a finer $\Delta t$ must be taken (*i.e.* $\Delta t \sim \tau/1000$), since the Lorentz force and the change in wave vector $k$ of a carrier in $\Delta t$ become too large to draw precise trajectories of carriers. The $z$-component of magnetoresistivity tensor $\rho_{zz}$ is calculated as:

$$\rho_{zz} = \frac{\sigma_{xx}\sigma_{yy} - \sigma_{xy}\sigma_{yx}}{\sigma_{zz}(\sigma_{xx}\sigma_{yy} - \sigma_{xy}\sigma_{yx}) - \sigma_{xz}\sigma_{yy}\sigma_{zx} + \sigma_{xy}\sigma_{yz}\sigma_{zx} + \sigma_{xz}\sigma_{yx}\sigma_{zy} - \sigma_{xx}\sigma_{yz}\sigma_{zy}} \qquad (2).$$

This $\rho_{zz}$ reduces to $1/\sigma_{zz}$, since the last three terms in the denominator are much smaller than $(\sigma_{xx}\sigma_{yy} - \sigma_{yx}\sigma_{yx})$ for (DMET)$_2$I$_3$. This is accurate to approximately one part in $10^3$.

## 5. Result and Discussion

The measured angle-dependent magnetoresistivity is shown in Fig. 1(b) in a 3D, polar surface plot. Presentation of the data in this manner reveals large oscillating features throughout all possible angular orientations of magnetic field. DKC oscillations are observed when rotating near $|\theta| = 90°$ at $\phi = 0°$ and $180°$. LMA oscillations are observed for $\theta$–rotations when $\phi = 90°$. The YAE is observed for a $\phi$ rotation with fixed $|\theta| = 90°$. Finally, the oscillations that appear for virtually all $\phi$ rotations with fixed $\theta$ are manifestations of the LN (or LNL) effect. Thus, all known AMRO effects are experimentally observed in the present single experiment. The calculated resistivity $\rho_{zz}(\theta,\phi)$ using the $\sigma_{ij}$ equation above is shown in Fig. 1(c), using the same polar format. Here, it can be seen that the calculated resistance is qualitatively and even semi-quantitatively in agreement with the experimental data, reproducing all known AMRO effects. This 3D representation of the experimental and calculated results is direct evidence of the multidimensional effect of magnetic field orientation on the transport properties of anisotropic materials. The relation between the several angular effects, if any, can be directly compared from a 3D visual presentation, something considerably less clear in conventional two dimensional $\rho_{zz}(\theta,\phi)$ plots, as were used in Ref. 16. A similar 3D plot of magnetoconductivity was previously reported in (TMTSF)$_2$PF$_6$ by Kang et al.,[40] showing the qualitative agreement between the experiment and calculations based on Ref. 33. However, the calculated magnetoconductivity doesn't exhibit oscillating features for the field rotation in pure *y-z* rotation (LMA orientation).

It is noteworthy that the dips observed in both experiment and calculation can be indexed by a generalized relation related to the real space lattice parameter of the materials;

$$\tan\theta\sin\phi = n\frac{a\sin\gamma}{c\sin\alpha\,\sin\beta^*} + \cot\beta^* \qquad (3),$$

where the integer *n* is the Lebed (or LNL) index, and $\cos\beta^* = (\cos\gamma\cos\alpha - \cos\beta)/(\sin\alpha\sin\gamma)$, using the crystal lattice angles $\alpha$, $\beta$ and $\gamma$. Also, it is seen that for the field parallel to the most conducting *x-y* plane ($|\theta|=90°$), the experimental magnetoresistance exhibits minima while the calculations show smooth variations with no such feature. Classically, for an isotropic 3D metal, the electron motion perpendicular to a magnetic field ($J \perp B$) experiences the largest Lorentz force, giving maximum magnetoresistance as shown by calculations. However, a broad dip in magnetoresistance is observed starting at $\phi \sim 15°$, with the deepest position at $\phi = 90°$. Similar behavior has been observed in some Q1D salts such as (TMTSF)$_2$PF$_6$ and (TMTSF)$_2$ClO$_4$ where the anomalous superconducting state existed with $H_{c2}$ far exceeding the Pauli limit, possibly associated with unconventional pairing [41,42,43]. However, $H_{c2}(T)$ for (DMET)$_2$I$_3$ tends to saturate at low temperature without exceeding the Pauli limit and it was also shown that the minimum observed in magnetoresistance is associated with a normal state field induced dimensional crossover, rather than superconductivity [44]. Strong *et al.* have suggested that a minimum can be observed at a large enough field parallel to this *y*-axis, which de-emphasizes coherent interlayer motion, transforming a 3D Fermi liquid into a 2D non-Fermi liquid [27]. Recently, based on experimental results on Q2D materials, the minimum for field parallel to the *x-y* plane was explained due to contributions from parallel coherent and incoherent transport [45]. To summarize, this 3D visualization of the experimental and calculated magnetoresistance shows evidence for the dimensional crossover of electronic motion due to magnetic field orientation. Like other theoretical models, the present calculation is unable to reproduce the experimentally-observed resistance minimum for field parallel to the *x-y* plane.

The first predicated and observed dimensionality effect (magic angle effect) is evident from a field rotation in the *y-z* plane, shown in Fig. 2(a). Also shown is our calculation, qualitatively in agreement with experiment, including a broad minimum around $\theta = 8.5°$ due to the triclinic crystal structure. The positions of higher order resistance minima coincide with the experimental data. In calculations, only lower order oscillations are clearly visible, with higher order oscillations evident in derivatives [16]. Furthermore, it is possible to estimate the three dimensional lattice parameters of these Q1D materials using magic angle effect.

DKC oscillations, observed for field rotations in the *x-z* plane, are compared with calculation in Fig. 2(b). These oscillations provide a way to measure the shape of the Fermi surface and the electronic band parameters to confirm the anisotropy of the material. A coherence peak is observed in both experiment and calculation, centered at $\theta = 90°$. The width of this peak is related to the strength of interlayer coupling, $t_z$ [9] compared to $t_x$ taking into account small electron closed orbits for field parallel to the *x*-axis via $\Delta\theta \approx 2t_z c/t_x b' \sin(b' k_F)$ [46]. Using $c = 1.5776$ nm, $b' = b/2 = 0.3881$ nm (*i.e.* ignoring a weak dimerization of DMET molecules along the 1D columns), $t_z/t_x = 1/300$, and $k_F = \pi/2b'$, the width of the coherence peak is calculated to be $\Delta\theta \sim 1.55°$. This value is in reasonable agreement with the experimentally observed width of $1.35°$. For field parallel to the x-axis, both open and closed orbits exist on the Fermi surface, with the fraction of closed orbits decreasing as $t_y/t_z$ increases [47]. A similar peak in magnetoresistance is also observed in Q2D conductors with warped Fermi surfaces [3]. Furthermore, two pronounced resistance peaks are observed around $\theta = (90\pm15)°$ that can be used to estimate the intraplane coupling, $t_y$ [9]. Using $c = 1.5776$ nm and $v_F = 4.0 \times 10^4$ m s$^{-1}$ [48], $t_y$ is estimated using the condition $J_0(\gamma_n) = 0$ where $J_0(\gamma_n)$ is Bessel function of zeroth-order and $\gamma_n = 2t_y c \tan\theta_n / \hbar v_F$ to be $t_y = 5.4$ meV [9]. This value is smaller than that estimated for the Bechgaard salt (TMTSF)$_2$ClO$_4$, $t_y = 12$ meV (24 meV above an anion ordering state at 24 K) [9] and for (TMTSF)PF$_6$ under a 10 kbar pressure [49]. $t_y = 32.5$ meV.

The YAE, observed for a magnetic field rotation in the x-y plane, is closely related to the corrugation of the Q1D Fermi surface within the plane [11]. In the YAE, a pair of resistance minima is observed, as shown in Fig. 2(c), which can be used to estimate the ratio of transfer integrals $t_x/t_y$ by using the following analytical expression [50], $\Delta\phi = 2\sqrt{2}\, y t_y \sin\gamma / x t_x$. With lattice parameters $x = b/2 = 0.388$ nm, $y = 0.6669$ nm, and $\gamma = 78.19°$ and observed angular width of $\Delta\phi \sim 30°$, the transfer integral ratio is estimated as $t_x/t_y \sim 9.0 \pm 0.9$. Since our experimental data are recorded every $5°$ in $\phi$, our $\Delta\phi$ estimation has an uncertainty which we estimate to be $\delta\phi = \pm 2.5°$. A somewhat more accurate estimation of $\Delta\phi$ can be found in the literature [38], which gives a value of $\Delta\phi = 28°$, corresponding to a transfer integral ratio $t_x/t_y = 9.7$. Furthermore, the same anisotropy of $t_x/t_y = 9.7$ is estimated from the superconducting anisotropic coherence length [44]. Thus, the estimated value of $t_x/t_y$ from the current measurement is in agreement with the previously reported ratio, within experimental uncertainty.

The more complex oscillations in Fig. 1 for every possible combination of the angle $\phi$ and $\theta$ are called LNL (originally named LN) for field rotating in an arbitrary plane. In Fig. 3(a), we compare calculation with experimental results for a magnetic field rotation off the *x-y* plane ($\theta \sim 82°$). The result is qualitatively in agreement with experimental data. These oscillations are very complex and their physical meaning has been elusive. Recently, Lebed and Naughton [29] proposed an "interference commensurate (IC)" nature related to special commensurate electron trajectories in a magnetic field, where an average electron velocity along the *z*-axis is non-zero. In particular, they demonstrated that, in the absence of Landau level quantization for open Fermi surfaces, this "other" (than Landau or Aharonov-Bohm) effect in a magnetic field, based on Bragg reflections, results in a series of 1D to 2D crossovers at the minima of the LN oscillations [30]. More to the point, these can be considered as 'multidimensional' crossovers, under various conditions involving 1, 2 and 3-dimensional electronic motion.

Figure 3(b) compares the field dependence of the magnetoresistance at the positions of the maxima and minima shown in Fig. 3(a), from both experiment and calculation. As predicted [29,30], $R_{zz}(B,\theta,\phi)$ tends to saturate at commensurate, 2D directions (minima), with nonsaturating, power-law behavior at incommensurate, 1D directions (maxima) [14]. The experimental data were also fit to a quadratic $B^2$ term, for comparison. Here, the basic idea of saturating magnetoresistance at commensurate angles (minima in angle sweeps) and non-trivial, non-saturation otherwise, is borne out in the experiments as well as in calculations. The crossover between 1D and 2D electron motion is the same phenomenon as was observed for (TMTSF)$_2$ClO$_4$ [14], thus demonstrating its general nature in "multidimensional" anisotropic metals.

## 6. Conclusion

We have presented measurements of the interlayer magnetoresistance of the Q1D organic molecular conductor (DMET)$_2$I$_3$ at low temperature and across all magnetic field orientations, where all AMRO effects are observed. We have numerically calculated the interlayer magnetoconductivity tensor for the same field orientations, using the triclinic lattice parameters. These AMRO are directly connected to the multidimensional electronic and crystal anisotropy of Q1D systems and are manifestations of the confined motion of electron motion under the influence of a magnetic field. The measurement of these oscillations is very useful to determine the shape of the Fermi surface (which has 1D, 2D and 3D characteristics), where measurement techniques that require Landau quantization are unavailable, as well as estimations of the lattice parameters.

**Acknowledgement**

This was support by the National Science Foundation, under Grant No. DMR-0605339.

**Figure Captions**

Fig. 1 (Color online) Three dimensional presentation of interlayer magnetoresistivity. (a) A schematic of a crystal showing various directions and angles. (b) The measured magnetoresistivity $\rho_{zz}(\theta, \phi)$ of $(DMET)_2I_3$ at 100 mK and 9 T. (c) The calculated 9 T magnetoresistivity ($T=0$) using Boltzmann transport equation and the true triclinic crystal symmetry of $(DMET)_2I_3$. Several angular features are observed in magnetoresistivity. The magnitude of the magnetoresistance at a given $\theta$ and $\phi$ is proportional to the radial distance from the origin.

Fig. 2 (Color online) Comparison of calculated (solid lines) AMRO effects with experimental results (dots). (a) Lebed angular effect for field rotation in *y-z* plane ($\theta$–rotation with $\phi = 90°$), (b) DKC effect for field rotation in *x-z* plane ($\theta$–rotation with $\phi = 0°$), and (c) YAE for field rotation in *x-y* plane ($\phi$–rotation with $\theta = 90°$).

Fig. 3 (Color online) (a) Angular-dependent magnetoresistance calculated for $\theta \simeq 82°$ compared with experimental data at 100 mK and 6 T. (b) Field dependence of magnetoresistance at commensurate and non-commensurate orientations as indicated by arrows in (a). For maxima, the magnetoresistance is nonsaturating in field (1D behavior), and for minima, it tends to saturation (2D behavior). The dotted line is a polynomial fit with $R_{zz}(B,\theta,\phi) \sim B^2$.

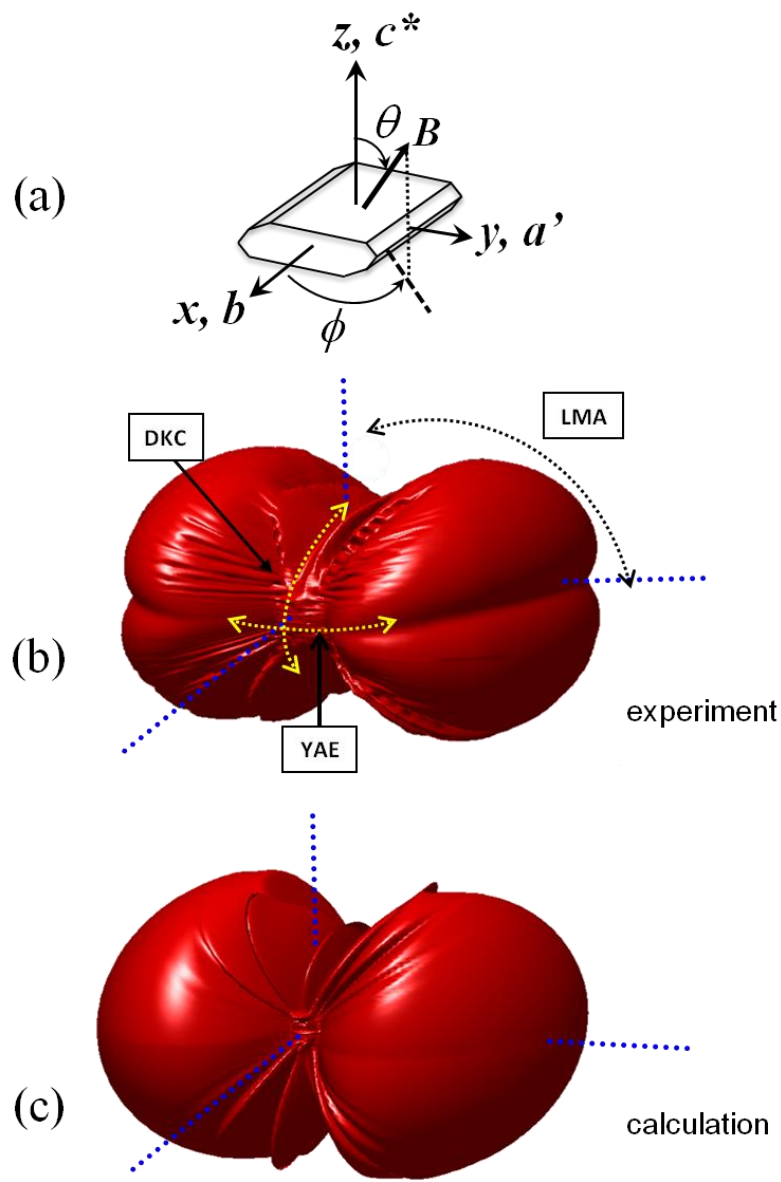

FIG. 1

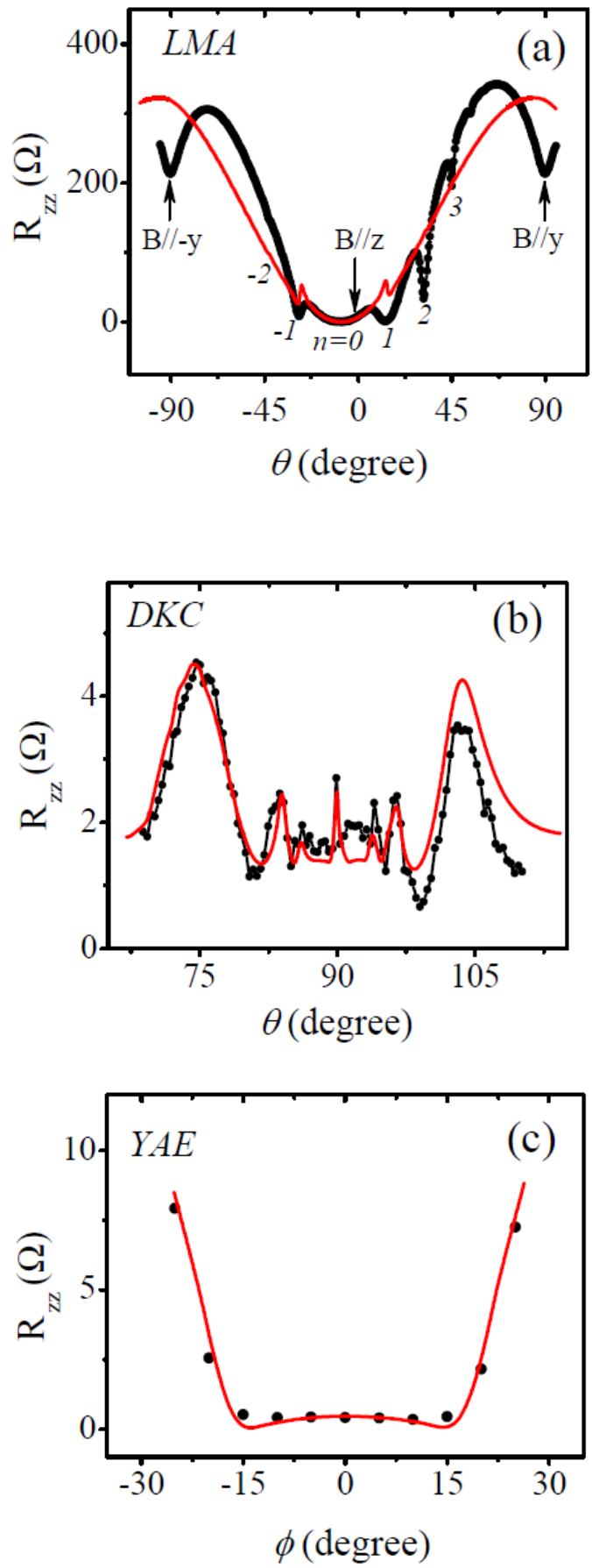

FIG. 2

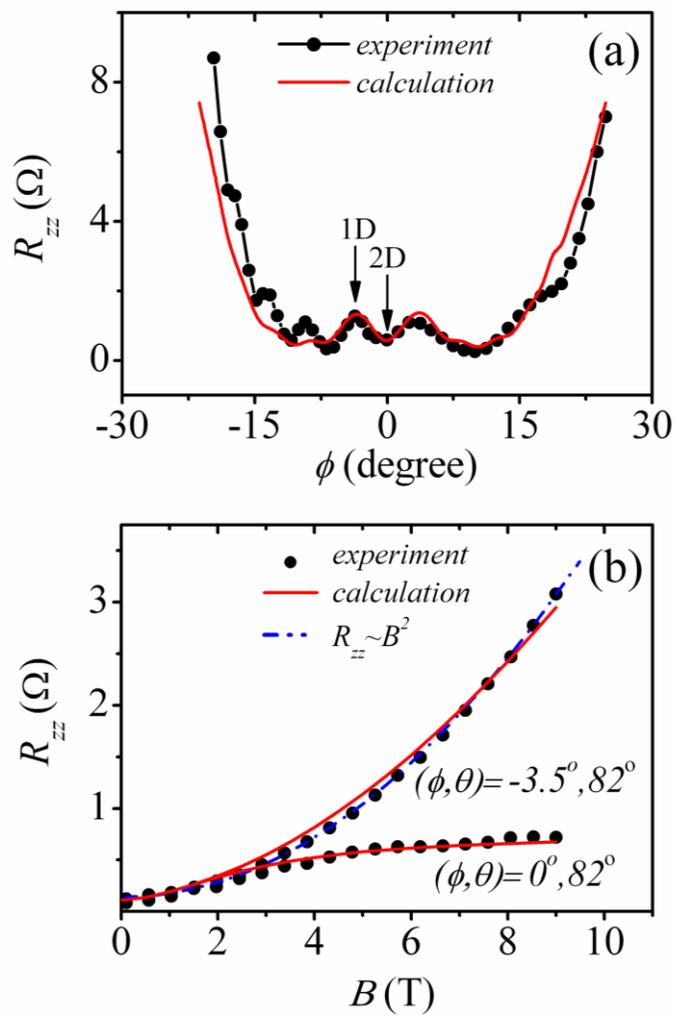

FIG. 3